\documentclass[manuscript]{aastex}

\shorttitle{Scatter-free pickup ions beyond the heliopause}
\shortauthors{S.V. Chalov et al.}

\begin{document}


\title{Scatter-free pickup ions beyond the heliopause as a model for
the Interstellar Boundary Explorer (IBEX) ribbon}


\author{ S.V. Chalov\altaffilmark{1}, D.B. Alexashov\altaffilmark{1,2},
D. McComas\altaffilmark{3,4}, V.V. Izmodenov\altaffilmark{1,2,5},
Yu.G. Malama\altaffilmark{1,2}, N.~Schwadron\altaffilmark{6,3}
\altaffiltext{1}{Institute for Problems in Mechanics Russian
Academy of Sciences} \altaffiltext{2}{Space Research Institute
(IKI) Russian Academy of Sciences} \altaffiltext{3}{Southwest
Research Institute, San Antonio, TX} \altaffiltext{4}{University
of Texas, San Antonio, TX} \altaffiltext{5}{Lomonosov Moscow State
University, Department of Aeromechanics and Gas Dynamics, Faculty
of Mechanics and Mathematics, Moscow, 119899, Russia;
izmod@ipmnet.ru} \altaffiltext{6}{Boston University, Boston, MA}
 }

\begin{abstract}
We present new kinetic-gasdynamic model of the solar wind
interaction with the local interstellar medium. The model
incorporates several processes suggested by \cite{mccomas09} for
the origin of the heliospheric ENA ribbon -- the most prominent
feature seen in the all sky maps of heliospheric ENAs discovered
by the Interstellar Boundary Explorer (IBEX). The ribbon is a
region of enhanced fluxes of ENAs crossing almost the entire sky.
Soon after the ribbon's discovery it was realized (McComas et al.,
2009) that the enhancement of the fluxes could be in the
directions where the radial component of the interstellar magnetic
field around the heliopause is close to zero (Schwadron et al.,
2009). Our model includes secondary charge exchange of the
interstellar H atoms with the interstellar pickup protons outside
the heliopause and is a further advancement of the
kinetic-gasdynamic model by \cite{malama06} where pickup protons
were treated as a separate kinetic component. \cite{izmod09} have
shown in the frame of Malama's model that the interstellar pickup
protons outside the heliopause maybe a significant source of ENAs
at energies above 1 keV. The difference between the current work
and that of \cite{izmod09} is in the assumption of no-scattering
for newly created pickup protons outside the heliopause. In this
limit the model produces a feature qualitatively similar to the
ribbon observed by IBEX.
\end{abstract}



\section{Introduction}

The collision of the supersonic solar wind with the interstellar
plasma flow results in formation of a complex interaction region
or heliospheric interface. This region includes the termination
and, possibly, bow shocks decelerating the solar wind (SW) and
interstellar plasma, respectively, and the heliopause separating
the two plasmas. The region of the heated SW behind the
termination shock (TS) is known as the inner heliosheath, while
the region behind the heliopause is called the outer heliosheath.
The local interstellar medium (LISM) is a partly ionized medium
consisting mainly of neutral atoms. It has become evident within
recent years that the interstellar atoms have a pronounced effect
on the global structure of the interface region and on the
physical processes operating in the heliosphere. Apart from the
fact that the position and shape of the TS and heliopause are
significantly determined by the action of the atoms, they give
rise to a specific hot population of pickup ions (PUIs). The first
direct measurements of pickup helium \citep{moebius85} and pickup
hydrogen \citep{gloeckler93} showed that the velocity
distributions of the PUIs differ in significant ways from the
velocity distributions of primary solar wind ions.

The first measurements of the IBEX (Interstellar Boundary
Explorer) spacecraft \citep{mccomas09, fuselier09, funsten09,
schwadr09} show results that were entirely unexpected. The
objective of the IBEX mission is to image the complex interaction
between the local interstellar medium (LISM) and the outflowing
solar wind by measuring the fluxes of energetic neutral atoms
(ENAs) originating in the outer parts of our heliosphere and
beyond. The first scan of the whole sky showed that maxima of ENA
fluxes form a long ($\sim 250-300^{\circ}$) and narrow ribbon-like
feature that was not predicted by any model prior to the IBEX
observations.

The speed of the original interstellar atoms entering the
heliosphere is $\sim$26.4 km/s, which for hydrogen atoms
corresponds to the kinetic energy of about 3 eV. Some portion of
the atoms experiences charge exchange with shock heated solar wind
protons and PUIs, and a new population of energetic atoms, created
as a result of this process, has the broad energy distribution
extending over several keV. These ENAs represent the energy
distributions of the parent charged particles and, therefore, when
measured at the Earth's orbit, can be used as a remote sensing of
the ions in the interaction region.

Current theoretical models of the SW/LISM interaction fall into
two categories: standard models which assume instantaneous
assimilation of pickup ions in the SW \citep{baranov93}, and the
compound or multi-component model \citep{malama06} in the
framework of which the pickup particles are considered as separate
isotropic (in the solar wind rest frame) populations with their
specific energy distributions. \cite{izmod09} presented an
extension of the \cite{malama06} model by introducing a
non-thermal population of pickup protons in the interstellar
medium. These authors showed that the interstellar pickup protons
form significant fluxes of ENAs dominating at energies above
$\simeq 1$ keV. Although the multi-component models are more
comprehensive, all of the current numerical models predict that
the ENA fluxes have maxima near the upwind direction of the
heliosphere and minima at the flanks, though, of course, the
position of maxima can slightly deviate from the upwind direction
due to effects of the interstellar magnetic field and solar wind
asymmetry (e.g. Izmodenov et al. 2009).

\cite{mccomas09} presented six possible concepts for the formation
of the ribbon observed by IBEX. Among these concepts was the idea
that neutralized solar wind propagates out beyond the heliopause,
becomes ionized, gyrates about interstellar magnetic field lines,
and then charge exchanges again to become ENAs. Some of these ENAs
move back in toward the Sun where they can be imaged by IBEX. The
advantage of this mechanism is that it produces sharply peaked ENA
emissions in directions roughly perpendicular to the interstellar
magnetic field beyond the heliopause - the same alignment inferred
by comparing the IBEX ribbon to an MHD simulation of the
heliosphere (Schwadron et al., 2009). Another concept suggested by
\cite{mccomas09} is that compression of the interstellar magnetic
field beyond the heliopause may cause ions to align preferentially
perpendicular to the interstellar magnetic field through
conservation of the first adiabatic invariant and conservation of
energy. This will also lead to a special orientation of peaked ENA
emissions perpendicular to the interstellar magnetic field, and
may help to explain both how the ribbon is formed and the even
more surprising fine structure observed in it (McComas et al.,
2009). The basic idea of secondary ENA generation of the IBEX
ribbon was further examined by \cite{heerick10}. These authors
assumed that pickup protons in the outer heliosheath have and
retain a partial shell distribution and that their
re-neutralization is effectively instantaneous. This approach is
significantly different from ours in this study since we solve
consistently for the motion of pickup protons along magnetic field
lines in the scatter-free limit, and thus include the motion of
PUIs along the field line between their pickup and
reneutralization.

\begin{figure}[t]
\noindent\includegraphics[width=7cm]{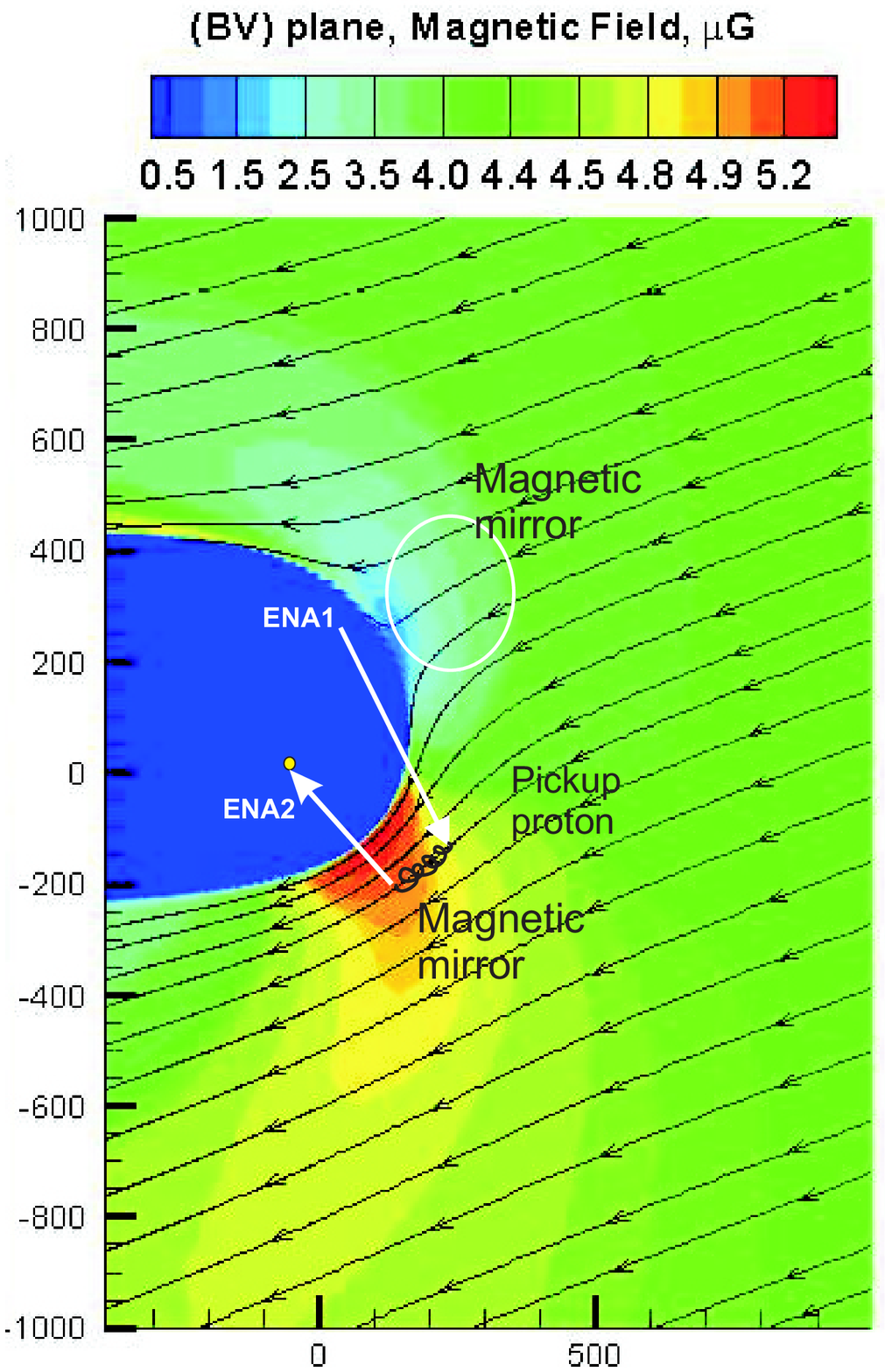} \caption{The
spatial distribution of the interstellar magnetic field around the
heliospause in the $\mathbf{BV}$ plane. The arrows show direction
of the magnetic field, while the color indicates the magnetic
field magnitude. The angle between $\mathbf{B}$ and $\mathbf{V}$
far from the heliopause equals $20^{\circ}$ and the magnitude of
$B$ is 4.4 $\mu G$.} \label{fig1}
\end{figure}

\section{Approaches to the problem}

The IBEX observations of an unexpected narrow ribbon of enhanced
ENA fluxes raise fundamental questions about the origin of these
particles. \cite{mccomas09} considered 6 possible sources of the
ribbon including both sources inside and beyond the heliopause.
The idea that the source lies in the outer heliosheath has at
least two arguments. First, it was recently demonstrated in
\citet{izmod09} that high-energy charged (pickup) protons can
arise in the outer heliosheath due to charge exchange between
interstellar protons and ENAs originated inside the heliopause.
The energy distribution of the pickup protons has maximum near 1
keV and extends to energies of about 10 keV. Secondly, it was
noted in \citet{mccomas09} and \citet{schwadr09} that the ribbon
position, as seen from the Earth, coincides closely with the
likely magnetic field direction located just beyond the
heliopause, where, according to the recent MHD models, the
interplanetary magnetic field is perpendicular to the heliocentric
radial direction. The latter circumstance means that the
interstellar magnetic field beyond the heliopause\footnote{Note
that neutrals from the solar wind will be picked up over a huge
range of distances from the heliopause ($\sim$1000 AU).} plays a
very important role in the formation of the ribbon. This role is
twofold. On the one hand, the dynamical effect of the magnetic
field essentially changes the shape of the heliosphere and the
pattern of the plasma flows in the interface region (see, e.g.
Izmodenov 2009). On the other hand, the magnetic field influences
the transport of energetic charged particles (PUIs). While the
primary interstellar plasma can be considered as a collisional
medium and can be described in the framework of the MHD approach
\citep{baranov00}, pickup protons originating in the outer
heliosheath from heliospheric ENAs with energies of about 1 keV
are collisionless. The more comprehensive global theoretical
models of the heliospheric interface \citep{malama06, izmod09},
treated PUIs as a separate population of charged particles, assume
that the velocity distributions of the PUIs in both inner and
outer heliosheathes are isotropic (in the plasma rest frame). In
other words, the isotropization time in these models is considered
to be the smallest characteristic time. This is fairly good
approximation for the supersonic solar wind and, possibly, for the
inner heliosheath. However, in the interstellar medium this time
is unknown. Here we consider the opposite limiting case when the
scattering of PUIs in the outer heliosheath due to wave-particle
interactions is completely ignored. We show that in this limiting
case a feature arises from simulations that is qualitatively
similar to the observed ENA ribbon, much as \cite{heerick10}
found.

\begin{figure}[t]
\noindent\includegraphics[width=8cm]{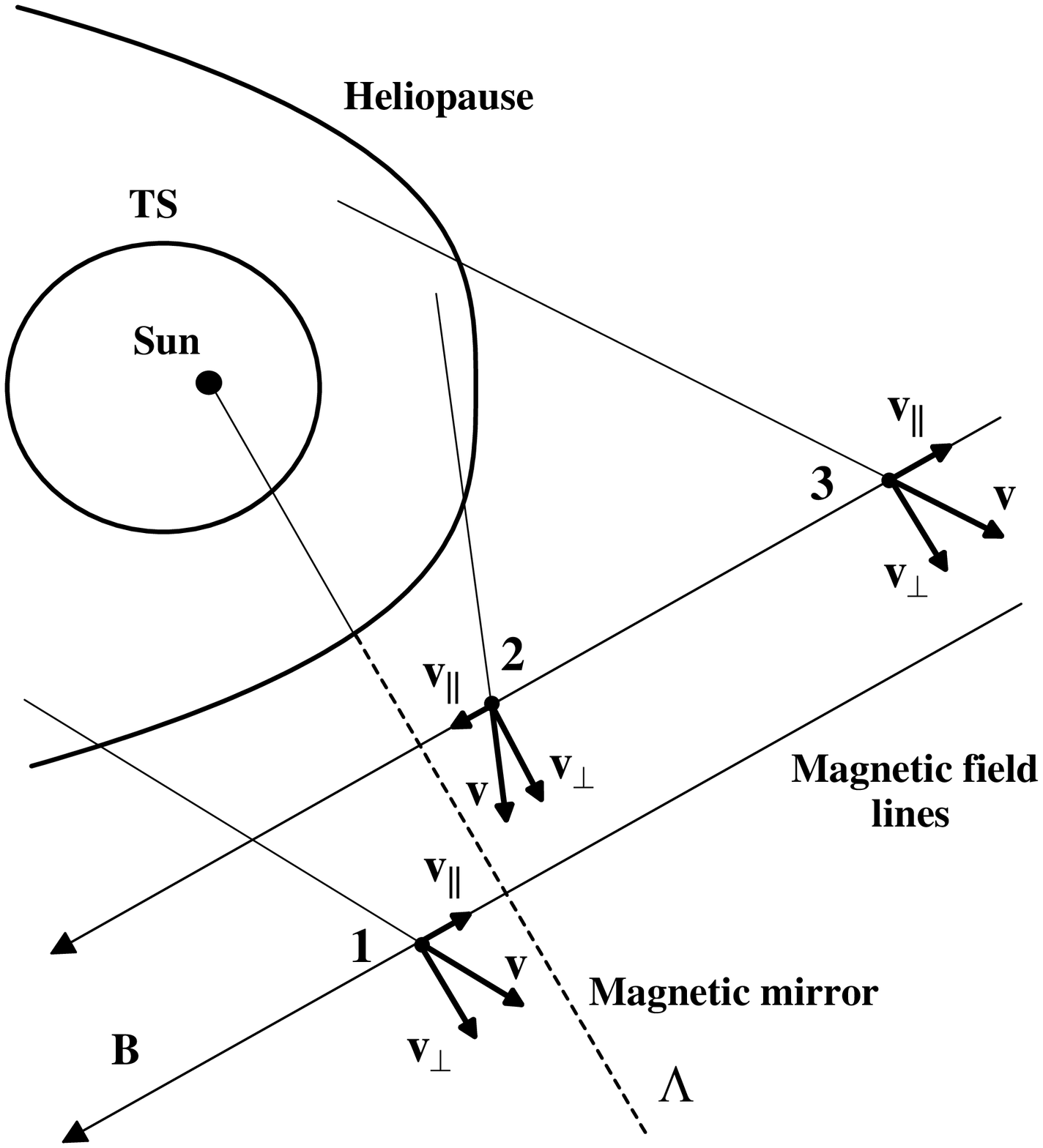} \caption{Pick-up
protons 1, 2 and 3 are born in the outer heliosheath owing to
ionization of heliospheric ENAs. Pick-up protons 1 and 2 can be
reflected at the magnetic field maxima (see also Fig.~\ref{fig1}),
while proton 3 moves away from the heliosphere.} \label{fig2}
\end{figure}

The population of pickup protons beyond the heliopause is the
product of the charge exchange process between interstellar
protons and heliospheric ENAs. The ENAs can be subdivided into two
type. Type 1 originates in the supersonic solar wind -- the
so-called neutral solar wind. The energy of these atoms is about 1
keV. Type 2 originates in the inner heliosheath. The ENAs from
this region have a broad energy distribution extending up to
several tens of keV.

\section{Numerical model}

In the case of negligible scattering, which we consider here, the
motion of a pickup proton in the outer heliosheath consists of
gyration around a magnetic field line and free motion along this
line. Figure~\ref{fig1} shows the spatial distribution of the
interstellar magnetic field around the heliopause in the
$\mathbf{BV}$ plane, where $\mathbf{B}$ is the magnetic field
vector and $\mathbf{V}$ is the vector of the interstellar plasma
velocity. The arrows show direction of the magnetic field, while
the color indicates the magnetic field magnitude. The results
presented here are obtained in the framework of the numerical
three-dimensional model, with an MHD description of the plasma
flows and a kinetic description of atoms \citep{izmod05}. The
process of generating an ENA in the outer heliosheath is also
shown schematically in Fig.~\ref{fig1}. An energetic atom from the
heliosphere (ENA1) penetrates into the outer heliosheath. Due to
the charge exchange reaction between the ENA1 and an interstellar
proton, a new pickup proton is "born". Once produced, it moves
along a magnetic field line until a subsequent charge exchange
results in the formation of a new energetic atom (ENA2). Under
appropriate conditions this new ENA reaches the vicinity close to
the Sun where it can be detected by IBEX.

In Fig.~\ref{fig1} one can see the domains of the increased
magnetic field magnitude and domains where the magnitude reaches
its minimal values. The transport of pickup protons in the outer
heliosheath is substantially determined by these features. The
regions of strong magnetic field can be considered as magnetic
mirrors or stagnation regions where the motion of charged
particles along field lines is decelerated and some portion of the
particles is reflected. Figure~\ref{fig2} schematically
illustrates the velocities of individual particles in the vicinity
of the magnetic field maximum presented in Fig.~\ref{fig1}. The
maximum of the magnetic field (the magnetic mirror) is marked by
the dotted line $\Lambda$. The velocity, $\mathbf{v}$, of a pickup
proton originating from a heliospheric energetic atom can be
obtained as the sum of velocity along the magnetic field line,
$\mathbf{v}_\parallel$, and the gyration velocity,
$\mathbf{v}_\perp$. In the case of no scattering, the magnetic
moment of charged particles propagating in the slowly varying
magnetic field is an adiabatic invariant; we use this
simplification in our model. Furthermore, we ignore the effects of
the drift motion of the pickup protons in the charge exchange
process. This assumption is well-founded since the speeds of the
drift motion are about 20 km s$^{-1}$, small compared with the
proper speeds of the particles. Thus we have:
\begin{equation}
v_\perp^2 /B = \textrm{const} \,, \qquad v_\perp^2 + v_\parallel^2
= \textrm{const} \,. \label{f1}
\end{equation}
Equations (\ref{f1}) determine the motion of a pickup proton in
the fixed magnetic field $\mathbf{B}$. Particles that move along
the field line in the direction of increasing field magnitude
(protons 1 and 2 in Fig.~\ref{fig2}), $v_\parallel$ decreases,
while $v_\perp$ increases due to conservation of the first
adiabatic invariant and conservation of energy. For some of
particles $v_\parallel$ may become zero in the region of the
increased magnetic field and then these particles are reflected.
This is magnetic mirror effect. In any case the parallel
velocities of pickup protons near the maxima of the magnetic field
magnitude are small, so that the pickup protons spend a
comparatively long time in these regions. Therefore, these regions
in the outer heliosheath are ideal places for production of ENAs.
Note that, as it can be seen in Fig.~\ref{fig1}, the radial
component of the magnetic field at the maximum equals zero. In
other words, the position of these regions coincides with the
observed position of the IBEX ribbon in the sky.

Many magnetic mirrors can exist in the vicinity of the heliopause.
In Fig.~\ref{fig1} we show two of them. In this way a charged
particle can be trapped between two mirrors until the charge
exchange reaction results in the formation of a ENA.

\begin{figure}[t]
\noindent\includegraphics[width=8cm]{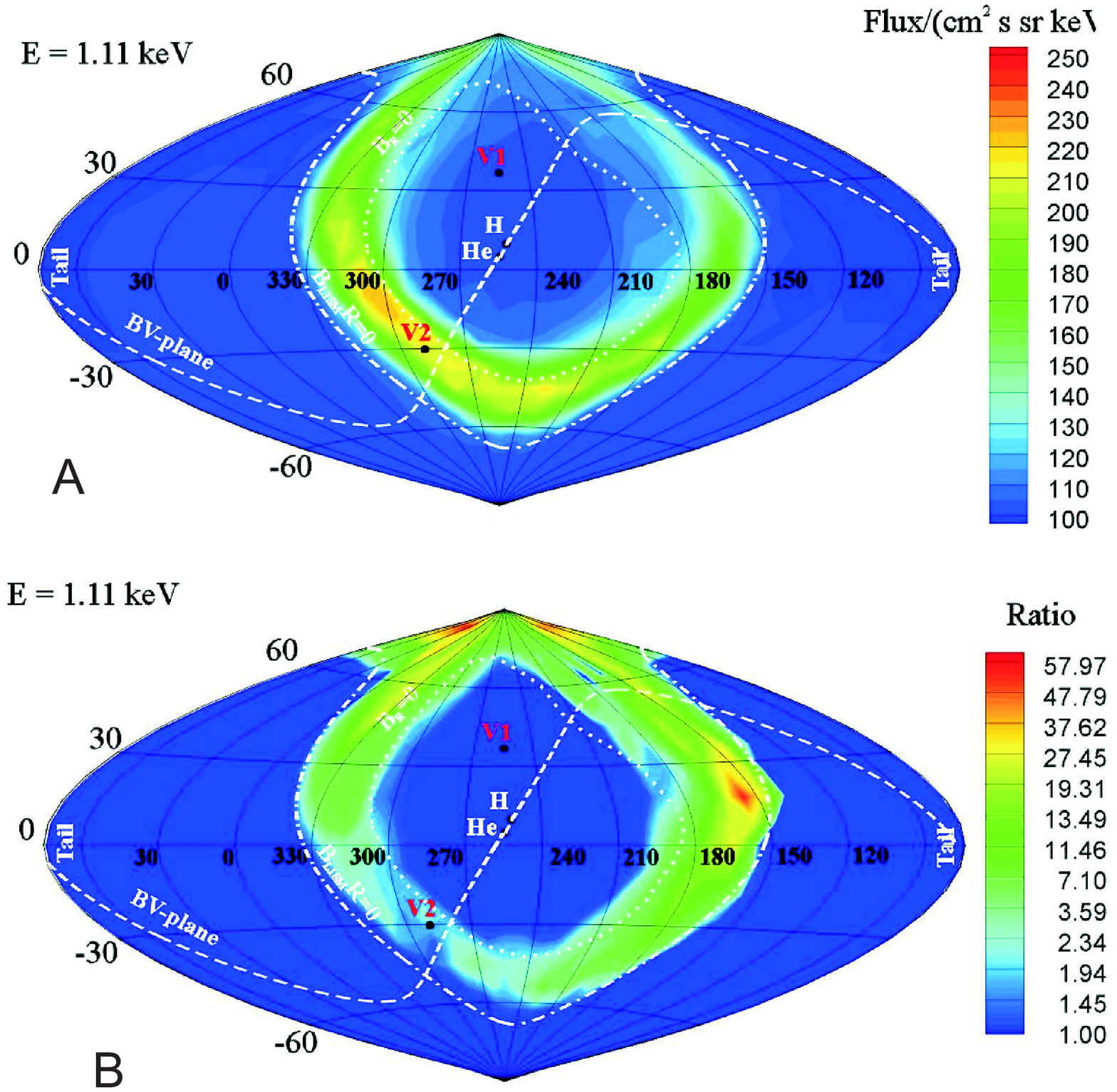} \caption{All-sky
maps of calculated ENAs in the energy range from 0.83 to 1.39 keV
in Mollweide projection in ecliptic coordinates. A) Differential
fluxes in (cm$^{2}$ s sr keV)$^{-1}$. The curves
$B_{\mathrm{R}}=0$ near the heliopause,
$\mathbf{B}_{\mathrm{LISM}} \mathbf{R}=0$ and the $\mathbf{BV}$
plane are shown. B) The ratio of ENA fluxes from the outer
heliosheath to ENA fluxes from the inner heliosheath in the case
of the Maxwellian distribution of the mixture of pickup and solar
wind protons. ENAs of sort 1 are excluded as parent atoms for
pickup protons in the outer heliosheath (see text).} \label{fig3}
\end{figure}

Results of our calculations of fluxes of energetic hydrogen atoms
at 1 AU from the outer heliosheath at the energy about 1 keV are
shown in Fig.~\ref{fig3}. The numerical model makes use of the
simplified guiding center approach for pickup protons, which is
based on conservation of the magnetic moment and energy, and on
the magnetic mirror effects. The interstellar pickup protons are
calculated using the Monte Carlo method (a similar code used to
simulate H atoms). The ENA fluxes are calculated directly in the
Monte Carlo code. There is a significant difference between our
model and the model by \cite{heerick10}, which also attempts to
explain the ribbon-like feature considering pickup protons in the
case of weak scattering. However, they assume that pickup protons
in the outer heliosheath have a partial shell distribution and
that re-neutralization is instantaneous. In other words, the
motion of the pickup protons along the magnetic field lines is not
included.

The ribbon-like structure, similar the IBEX ribbon, is clearly
seen in Fig.~\ref{fig3}A. This figure shows the all-sky map of
calculated differential ENA fluxes in the energy range from 0.83
to 1.39 keV in Mollweide projection in ecliptic coordinates. We
note that our 3D model currently treats pickup and solar wind
protons in the inner heliosheath as a single fluid with a
Maxwellian velocity distribution (a 3D multi-component model is
under development). Thus the dominant contribution to the fluxes
in Fig.~\ref{fig3}A is due to pickup protons, originating from
heliospheric ENAs of type 1 (supersonic solar wind ENAs). Only
this population is responsible for the appearance of a ribbon-like
feature in our current simulations. Because fluxes of ENAs of type
2 from the inner heliosheath are low for the Maxwellian
distribution of the mixture of pickup and solar wind protons, we
artificially added background fluxes of 100 (cm$^{2}$ s sr
keV)$^{-1}$. In a forthcoming study based on a 3D multi-component
model, the background fluxes will be calculated self-consistently.

Our estimates show that the magnetic mirror effect producing
multiple reflections of particles is not very important for pickup
protons, originating from heliospheric ENAs of type 1. Taking this
effect into account produces a 25\% increase in ENA fluxes from
the outer heliosheath. However, the mirroring can be very
important for pickup protons, originating from heliospheric ENAs
of type 2 (from the inner heliosheath). Figure~\ref{fig3}B shows
the calculated ratio of ENA fluxes from the outer heliosheath to
ENA fluxes from the inner heliosheath for a Maxwellian
distribution approximating the mixture of pickup and solar wind
protons. The ENAs in the outer heliosheath arise as a result of
charge exchange between interstellar atoms and pickup protons,
which, in turn, are produced due to ionization of ENAs of type 2
from the inner heliosheath. Thus, ENAs of type 1 (supersonic solar
wind ENAs) in Fig.~\ref{fig3}B are excluded as parent atoms for
pickup protons in the outer heliosheath. Since, in reality, the
velocity distributions of protons (solar wind + pickup) in the
inner heliosheath are far from Maxwellian and have substantial
suprathermal tails, we consider the current results as only
illustrative of the mirroring and accumulation effects of pickup
protons in the nonuniform magnetic field, which is schematically
shown in Fig.~\ref{fig2}. Figure~\ref{fig3}B shows that these
effects result in the formation of a pronounced ribbon-like
structure even with a Maxwellian velocity distribution in the
inner heliosheath.

\section{Conclusions}

Here we have reported a new model without scattering, but
including the effects of ion transport for the pickup protons
generated in the region outside of the heliopause by charge
exchange of the thermal interstellar protons and heliospheric
ENAs. The results of the model yield a feature qualitatively
similar to the IBEX ribbon. In future studies the results of
simulations will be quantitatively compared to IBEX ENA
observations. These further studies need to take into account ENAs
for the inner heliosheath in proper kinetic way as it was done in
\cite{malama06}.

Acknowledgements. The work of S.C., D.A., Y.M. was supported in
part by Rosnauka under goskontrakt 02.740.11.5025 and by the RFBR
in the frames of the projects 08-02-91968-DFG, 10-01-00258,
10-02-01316 and the Program for Basic Researches of OEMMPU RAS.
V.I. was supported by President Grant MD-3890.2009.2 and Dynastia
Foundation. Work in the USA was supported by the IBEX mission,
which is a part of NASA's Explorer Program.

\end{document}